# Efficiency Assessment of a Residential DC Nanogrid with Low and High Distribution Voltages Using Realistic Data

Saeed Habibi, Ramin Rahimi, Pourya Shamsi, Mehdi Ferdowsi
Department of Electrical and Computer Engineering
Missouri University of Science and Technology
Rolla, USA
s.habibi@mst.edu, r.rahimi@mst.edu, shamsip@mst.edu, ferdowsi@mst.edu

*Abstract*— **Direct Current (DC) power distribution has gained attention in the Residential Nanogrids (RNGs) due to the substantial increase in the number of roof-top Photovoltaic (PV) systems and internally DC appliances used in buildings. Using DC distribution improves the efficiency of the RNGs compared to AC distribution. This paper investigated the efficiency of a DC RNG for low and high distribution voltage levels by exploring reasons for power losses. The studied DC RNG consisted of various types of local loads, on-site PV generation, and battery storage systems. The realistic load, PV profiles, and converter efficiency curves were used to make the analysis more accurate. In addition, three load profiles with low, medium, and high power consumptions were considered to study the load impacts on the overall system efficiency.**

## I. Introduction

Nowadays, due to the continuous rise in direct current (DC) compatible loads, photovoltaic (PV) system installations, and battery storage use in residential and commercial buildings, considerable attention has been given to the implementation of DC distribution systems in residential nanogrids (RNGs) [1]. DC distribution systems are more efficient for data center applications than the AC counterpart [2]. Although, unlike data centers, less energy is saved in residential applications [3],[ 4], DC power distribution in RNGs provides high efficiency and cost savings [5]. The performance and efficiency of DC distribution for residential and commercial buildings have been investigated for the last two decades [5]-[10]. In DC powered residences total electricity savings of 14% and 5% were estimated with and without energy storage [5]. In [6], a detailed model was used to compare equivalent AC and DC building distribution networks with PV generation and battery storage. According to [6], a medium office building had 18.5% energy savings using a DC distribution system, compared to using an AC distribution system.

Due to lack of standardization, previous research on residential and commercial DC buildings offered various DC voltage levels for DC distribution ranging from 12 V to 400 V DC [7], [8]. In [9], power losses and performance of a DC stand-alone nanogrid were studied at low voltages from 12V to 100V, and this paper concluded that 48V voltage was the best choice for low voltage DC distribution for a house. [10] studied 48V, 220V, and 380V DC and compared it with the 220V AC. However, in these studies, variations in load, PV generation, and converters efficiency were not involved.

This study focused on the efficiency assessment of a DC RNG and analyzed the impact of DC voltage level, battery storage, and load on a DC RNG, incorporating both wiring losses and power conversion losses. This study was developed using realistic data for PV generation, loads, and converters. Using realistic data included the variations of PV generation, loads, and converters' efficiency into the study.

In this paper, an RNG was analyzed for two DC voltage levels of 48 V and 220 V, which were considered the low and the high distribution voltage levels, respectively. One reason that DC RNGs are appealing is they use on-site PV generation efficiently [11]-[13]. A PV plant power generation data, which is geographically close to the load model's location in Florida, was used [14]. In addition, DC RNGs can benefit from local battery energy storage systems. Due to their inherently DC nature, compared to AC nanogrids, fewer conversion stages are required for charging and discharging [15]. In this study, the impacts of adding battery storage to DC RNGs also were investigated through simulations.

Three different load models based on the Building America B10 benchmark are considered. These data were measured in Orlando and provided by the US Department of Energy (DOE) [12]. These three load models represent low, medium, and high power residential loads. Simulations were done for each load model at both low and high voltages.

## II. Nanogrid Structure and Components

A DC distribution system for the RNG, as shown in Fig. 1, was used for this study. This DC RNG utilized on-site PV generation and batteries. Based on the residential load data [12], loads are categorized into four sections: Heat-Ventilation-air conditioning (HVAC), Lights, Water Heater, and Interior Equipment (IE).

PV panels and batteries were connected to the DC-bus using DC-DC converters. The DC-DC converter connected to PV panels provides maximum power point tracking (MPPT) and also converts varying voltage from the PV panels to a fixed voltage [17], [18]. The battery converter is a bidirectional DC-DC converter that implements charging and discharging algorithms. A bidirectional AC-DC converter was used as an interface converter between the DC RNG and AC grid. The direction of power flow in the RNG is shown in Fig. 1 with arrows.

Loads were assumed to be inherently DC, which meant they could be connected directly to the DC bus. Based on the survey done on DC-ready appliances [19], most of these appliances are use 48 V. Additionally, the conventional appliances —with input rectification stage or variable frequency loads as in HVDC or IE— are compatible with 220 V DC.

*A. Loads*

In this study, three load models based on the DOE residential load data were selected, including low load, base load, and high load models. These load model data were measured in three different houses during a year with a one-hour sampling time for each load model. The low, base, and high load models represented small, medium, and large size houses, respectively. Each of these load models used four sets of data as categorized in Fig. 1.

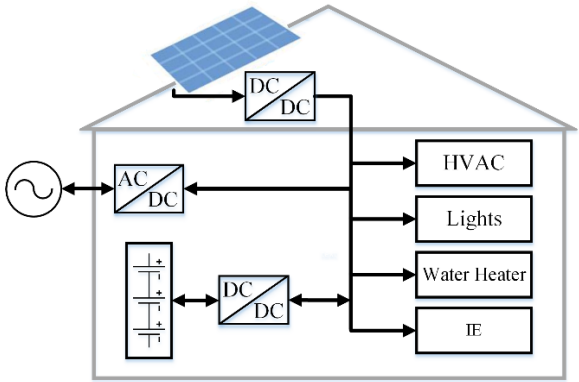

Fig. 1. Nanogrid structure.

*B. Converters*

The small number of power conversion stages in a DC RNG, increases efficiency. In this study, only three converters were required: a bidirectional AC-DC converter as an interface converter between AC grid and DC RNG, a bidirectional DC-DC battery charger converter, and a DC-DC PV converter. The efficiency curve of the converters were used for simulations to provide accurate and reliable results. Since some of the required converters may not have been commercialized, converters proposed in papers were used for simulation. In this study, the efficiency curves of XTRA3415N, LCM3000W-T, and HPQ-12/25-D48 were used for the PV, AC-DC, and battery converter at 48 V, respectively. At 220V, the proposed converters in [20], [21], and [22] were used for the PV, AC-DC, and battery converter, respectively. The efficiency curve of these converters at 48 V and 220 V are shown in Fig. 2. At every conversion stage, the DC RNG was assumed to contain enough parallel converters to meet the peak power requirements.

*C. Floor paln and wiring resistance*

Based on the Building America B10 benchmark, a floor plan for the low load model with 94 $m^2$ floor area was considered. The floor plan and dimensions of the house parts are shown in Fig. 3. This floor plan was used to identify the wiring length for every item that required an electrical connection, and from the wiring diagram shown in Fig. 3, the wiring length for each item was calculated. For each category of loads, a circuit was considered and the wire cross-sectional area for each circuit in the house was identified using 150% maximum current passing through that circuit.

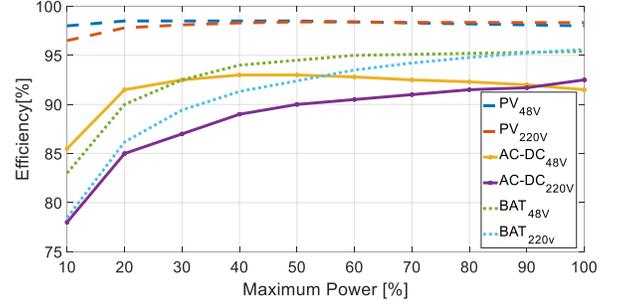

Fig. 2. Efficiency curve of converters.

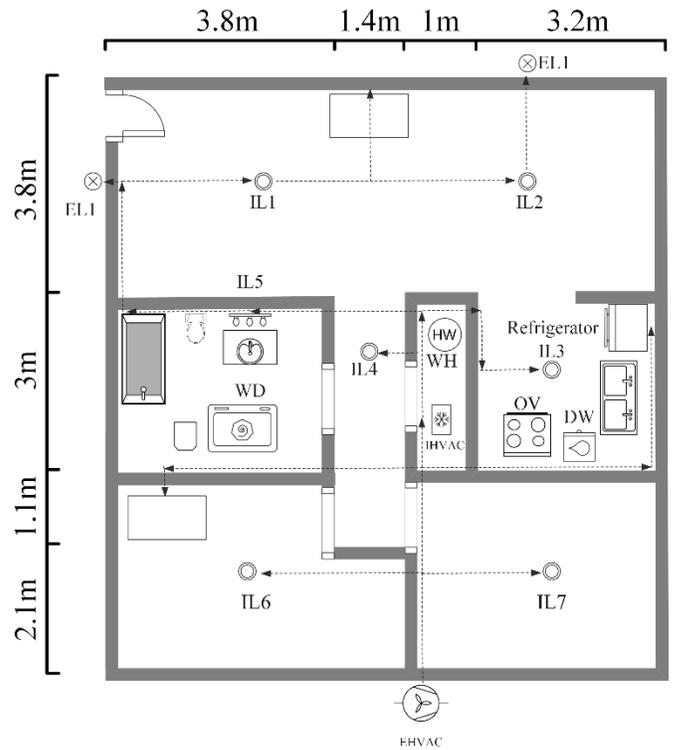

Fig. 3. Floor plan and wiring diagram.

Consequently, by using the wiring length and the wire cross-sectional area, the wiring resistance for each item can be calculated. Wiring loss for each category of loads can be calculated using (1). In (1), $R_i$ is the item resistance, $N$ is number of items in a specific category of load, and $I$ is total current of the specific category of load. For example, in lighting category, $R_i$ is the wiring resistance of each lamp, $N$ is number of the lamps, $I$ is total current of lighting load, and $P_{loss}$ is total lighting loss.

By applying the same procedure to each load category, equivalent wiring resistance for each category can be calculated using (2).

$$P_{loss} = \sum_{i=1}^{N} R_i \left(\frac{I}{N}\right)^2 = \sum_{i=1}^{N} \left(\frac{R_i}{N^2}\right) I^2 \quad (1)$$

$$R_{eq} = \sum_{i=1}^{N} \left(\frac{R_i}{N^2}\right) \quad (2)$$

Wiring lengths for the base load and high load models can be estimated based on Fig. 3. The floor areas of the base load and high load models were $188\ m^2$ and $282\ m^2$, respectively. Based on the floor area ratio, equivalent wiring resistances for the base and high load models were estimated as shown in Table 1. $R_{eq}$ was used in the simulation to calculate the wiring loss for each category.

Table 1. Equivalent wiring resistance for each catergory of loads.

| Req(mΩ) | Low Load | | Base Load | | High Load | |
|---|---|---|---|---|---|---|
| | 48 V | 220 V | 48 V | 220 V | 48 V | 220 V |
| HVAC | 21.35 | 21.35 | 4.71 | 30.18 | 2.28 | 23.23 |
| Lighting | 14.5 | 14.5 | 20.50 | 20.50 | 25.11 | 25.11 |
| IE | 13.2 | 13.2 | 18.66 | 18.66 | 22.86 | 22.86 |
| WH | 85.41 | 85.41 | 120.78 | 120.76 | 147.93 | 147.93 |

*D. PV*

The DC RNG used PV panels and an MPPT converter, which conditiones the panels' power and provides the maximum available power to the RNG. Different load models require different solar capacities[23]. In a zero net energy house, the solar capacity for each load model was obtained by matching the annual values of the generated and consumed energies, meaning that, the annual generated PV power should match the annual RNG's power consumption. In (3), $p_i^{pv}$ represents the PV plant power in [14], $p_i^{load}$ is the RNG's load, and $k$ is scaling factor for each load model. $k$ is determined by using (4), which means that annual load power and PV power are equal. Based on the calculations, solar capacities for the low, base, and high load models are 4kW, 8 kW, and 14 kW.

$$\sum_i p_i^{load} = \left(\frac{1}{k}\right) \sum_i p_i^{pv} \quad (3)$$

$$k = \frac{\sum_i p_i^{pv}}{\sum_i p_i^{load}} \quad (4)$$

*E. Battery*

A battery can be an energy source or sink when charging and discharging, respectively. Similar to solar capacity, battery energy storage capacity for each load model was chosen based on the annual load. Although there are various optimization approaches for determining battery size in stand-alone systems, there is not a specific method for grid-connected systems. In an ideal condition, the battery must be able to store the excess energy of a solar system and provide energy to the RNG when PV panels do not generate power. While it may result in a substantial unutilized capacity of the battery. Batteries are expensive, thus smaller capacity for the battery storage is desirable. However, a small battery capacity may lead to a deep discharge during non-daylight hours. When the load consumed more energy than the stored energy, state of charge (SOC) of the battery reached to the minimum allowable value and the battery controller disconnected the batteries from the RNG. The batteries stayed disconnected until PV generation exceeded load demand. Then the controller started charging the batteries. The period that the batteries stayed disconnected is defined as battery downtime (BDT). It is important to minimize BDT and optimize efficiency when evaluating battery capacity for a grid-connected system.

For the low load model, a simulation was performed by sweeping battery capacity from 2.4 kWh to 48 kWh to assess battery size effects on the efficiency and BDT. As the efficiency and BDT curves in Fig. 4 indicate, increasing the battery capacity reduced BDT rapidly at smaller capacities, but from approximately 20kWh, increasing capacity did not make a considerable change on BTD, and it decreased system efficiency. Therefore, battery capacity of approximately 20 kWh for the low load model was the best choice. A 19.2 kWh battery storage system was chosen for the low load model. Using the same battery sizing criteria for the base and high load models, 33.6 kWh, and 67.2 kWh battery storage systems were chosen for the base load and high load models, respectively.

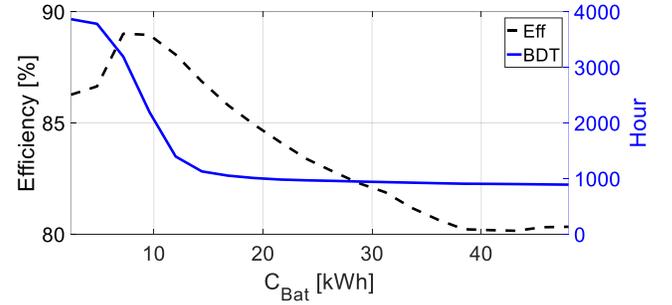

Fig. 4. System efficincy and battery down time.

The battery controller used in this study is a simple controller that only charges the batteries when PV excess power is positive and discharges the batteries when PV excess power is negative. The maximum allowable charging and discharging power was a quarter of the nominal battery capacity[17], and the state of charge (SOC) was limited between 100% to 20% for lead-acid batteries.

III. RESULTS

Simulations were performed on the low, base, and high load models at 48 V and 220 V distribution voltages. The simulation results helped to understand how the efficiency of the DC RNG varied from the low to the high DC distribution voltage. In addition, impacts of the load on the RNG's efficiency were investigated at low and high voltage levels. As the results shown in Fig. 5, simulations were done for the low load model ($LLM$), base load model ($BLM$), and high load model ($HLM$) at two voltage levels, which are indicated as subscripts. The efficiency of each load model has two bars, which represents the RNG's efficiency with and without batteries. Three phenomena can be observed from the results in Fig. 5:

- At each load model, the high voltage distribution had lower efficiency than the low voltage distribution.
- The efficiency of DC RNG slightly decreased as the load increased.
- Adding batteries to a DC RNG did not result in efficiency improvement.

By looking at the breakdown of the losses for each load model, the reasons behind the three phenomena were clarified. The breakdown of the losses for the $LLM$ is shown in Fig. 6. In this figure, losses are shown in four categories: PV inverter loss, AC-DC converter loss, battery converter loss, and wiring loss. Each bar represents the share of a specific category's loss to the total load of the RNG. The breakdown of losses for the $BLM$ and $HLM$ are shown in Fig. 7 and Fig. 8, respectively. The explanations for the occurrence of the three phenomena were described in the following section.

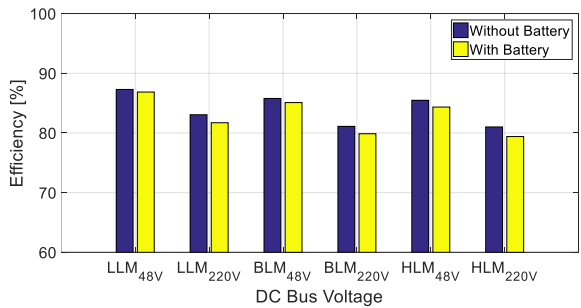

Fig. 5. Efficiency of the RNG with and without battery.

First, at each load model, the efficiency of the 220V RNG was lower than 48V RNG. For example, the efficiency of $LLM_{48V}$ was 87.29% while the efficiency of $LLM_{220V}$ was 83.78%, without batteries. As the breakdown of losses for the $LLM_{48V}$ in Fig. 5 indicates, 220V system had higher AC-DC and PV converter losses but lower wiring losses. The difference in PV converter loss between 48 V and 220 V RNG was negligible; furthermore, the wiring loss was less than 1% of total power at 48V and 220V RNG. Therefore, the main reason for the higher loss in 220 V RNG was that the AC-DC converter produced more loss at 220 V RNG. The higher loss was caused by the poorer efficiency curve of the converter at 220 V RNG. As shown in Fig. 2, the AC-DC converter at 220 V RNG had higher efficiency at its full load power, and it had lower efficiency within the powers less than 90% of its nominal power.

Second, as results in Fig. 5 indicate, the efficiency of each load model slightly decreased as the RNG's load increased. For example, 87.29%, 85.78%, and 85.47% were efficiency values for $LLM_{48V}$, $BLM_{48V}$, and $HLM_{48V}$, respectively. By comparing the loss breakdowns of the three load models, it was seen that while the AC-DC converter loss and the wiring loss increased, the power loss of the PV converter did not change considerably. 10.50%, 11.74%, and 11.98% were the AC-DC converter losses, and 0.45%, 0.7%, and 0.75% were the wiring losses for the $LLM_{48V}$, $BLM_{48V}$, and $HLM_{48V}$, respectively. It was concluded that higher load brought more wiring and AC-DC converter losses.

Third, as shown in Fig. 5, $LLM_{48V}$ efficiencies without battery and with battery were 87.2% and 86.8 %. It means that adding batteries to the DC RNG slightly reduced overall efficiency. Like $LLM_{48V}$, $LLM_{220V}$ had lower efficiency with batteries. As the efficiencies of the base and high load systems indicate, adding batteries to the Dc RNG, did not result in efficiency improvement.

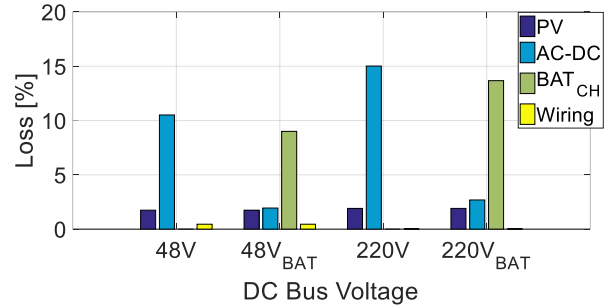

Fig. 6. Losses in the low load model.

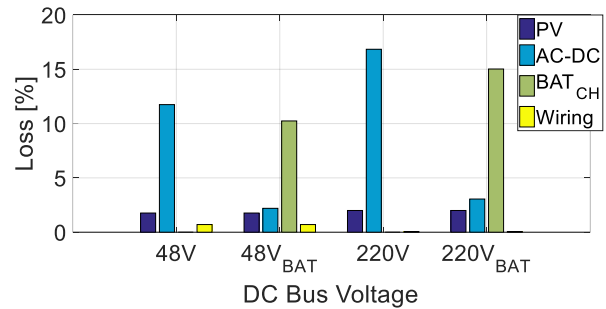

Fig. 7. Losses in the base load model.

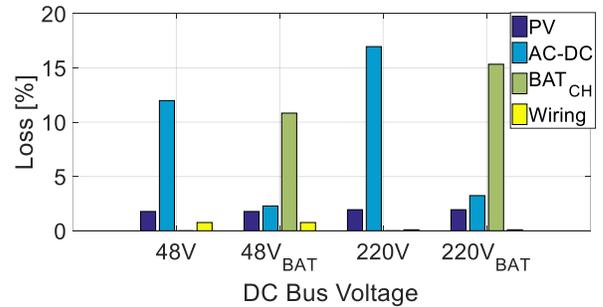

Fig. 8. Losses in the high load model.

By taking an in-depth look into the breakdown of losses in Fig. 6, it was noticed that by adding batteries to this system, the AC-DC converter loss dropped from 10.5 % to 1.94%, yet the battery converter loss increased from 0 to 9% of the total power. The PV converter loss and the wiring loss did not change in the presence of batteries. The drop in the AC-DC converter loss was due to a considerable reduction in the power flowing through this converter at the batteries' presence. Without batteries, the excess PV power flowed to the grid during days, and the grid supplied the load power during nights. Utilizing batteries in the RNG reduced the amount of power passing through the AC-DC converter by storing energy during days and supplying the load locally during nights. The converters' share of losses at $LLM_{220V}$, followed the same pattern as in $LLM_{48V}$. The wiring loss at $LLM_{220V}$ was less than 1%, therefore, there was not a bar representing the wiring loss. By comparing the breakdown of losses with and without batteries at BLM and HLM, the same pattern was seen in losses. Given these points, the AC-DC converter loss decreased significantly in all load models when batteries were added; however, high battery converter loss reduced system efficiency in the presence of batteries.

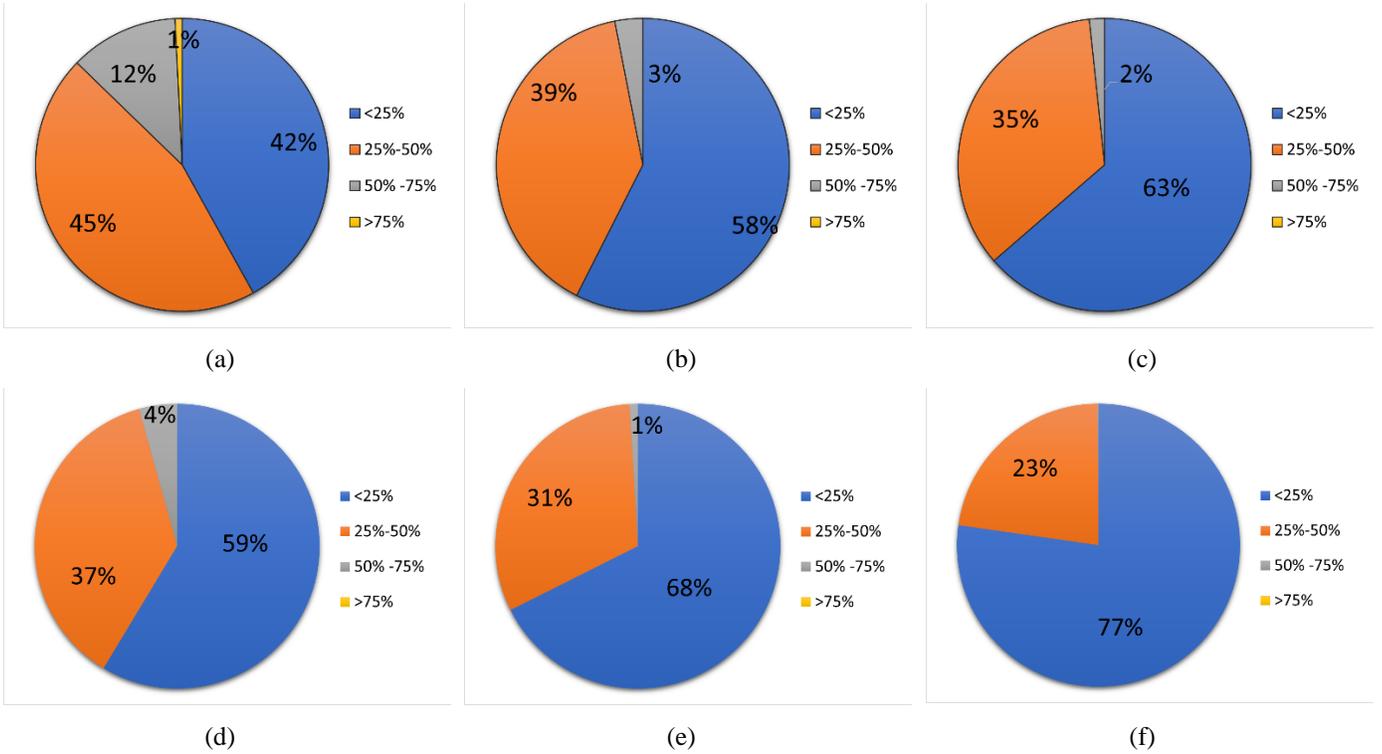

Fig. 9. Share of each operation region (a) AC-DC converter $LLM_{48V}$ without battery, (b) AC-DC converter $BLM_{48V}$ without battery, (c) AC-DC converter $HLM_{48V}$ without battery, (d) Battery converter $LLM_{48V}$, (e) Battery converter $BLM_{48V}$, (f) Battery converter $HLM_{48V}$.

As stated, the AC-DC converter was the main source of power loss in the DC RNG without batteries, and adding batteries to the DC RNG made the battery converter the main source of power loss, resulting in overall efficiency reduction. To understand the reason behind the efficiency reduction of the DC RNG with batteries, the operating region of the converters was considered. Fig. 9 (a) represents the share of each AC-DC converter operating region throughout a year. Converters' operating regions were divided into four quarters: less than 25%, between 25% to 50%, between 50% to 75%, and more than 75% of the converters' nominal power. It is desirable for converters to operate in their nominal power, but in practice, it is not possible. As shown in Fig. 9 (a), the AC-DC converter operated 42% of the time within the first quarter, 45% of the time within the second quarter, 12% of the time in the third quarter, and only 1% of the time within the fourth quarter of its nominal power at the $LLM_{48V}$ without batteries. Fig. 9(a) revealed that the AC-DC converter operated most of the time within the low-efficiency region, and because power was continuously transmitted between the grid and the RNG in a RNG without batteries, high power loss occurred in the AC-DC converter and ultimately reduced the efficiency of the RNG.

Similar to the AC-DC converter in $LLM_{48V}$, the share of operating regions for the battery converter is shown in Fig. 9 (d). The battery converter operated 59% of the time in the lowest power region, and 25% of its capacity remained unutilized during a year. It means that this converter was working in the lowest region of its efficiency curve most of the time and never utilized the high-efficiency region. This converter's low power operation brought a significant amount of undesired power loss to the RNG and eventually reduced the RNG's efficiency. A significant proportion of the RNG's power was processed by the battery converter, thus it was essential to use it in the high efficiency region. Otherwise, the efficiency of a DC RNG with batteries could not be greater than that of a DC RNG without batteries.

Comparison of the AC-DC and battery converters' share of the operating region in Fig. 9 revealed that the proportion of lower efficiency regions for the battery converter was greater than that of the AC-DC converter, and that is why the RNG with batteries had a lower efficiency than the RNG without batteries.

According to the results in Fig. 9 for the RNG without batteries, as the load increased from the low to the high, the low-efficiency region's share became dominant, which means that at higher loads, the AC-DC converter generated more power loss, which helps to understand the reason behind the second phenomena.

IV. CONCLUSION

This work aimed to assist industrial and academic decisions by assessing efficiencies between high voltage and low voltage DC distribution for an RNG. This work used a comprehensive RNG model based on real data to conduct simulations of the RNG at two voltage levels, with and without batteries. The results of simulations showed that the power loss of the converters contributed most to the system's overall loss. The AC-DC converter was the dominant source of loss among converters, while the addition of batteries to the RNG made the battery converter the dominant source of power loss.

By investigating the operating regions of the converters, it was found that as the load of the RNG increased, the AC-DC converter worked more within lower efficiency regions. Therefore, higher-load RNGs suffered greater losses. Moreover, the addition of batteries to the RNG increased overall loss. This phenomenon occurred because the battery converter operated for a significant time in the low power operating regions. It may appear that having a lower capacity for the battery converter would lower the power loss, but battery converter rating power was designed based on the maximum allowable charge and discharge rate, and since PV power generation and load power were not predictable, battery converter rating cannot be lower than the designed value.